\documentclass[aps,twocolumn]{revtex4-1}
\usepackage{subfig}

\usepackage{bbold}
\usepackage{amsmath}
\usepackage{multirow}
\usepackage{amsfonts}
\usepackage{bm}
\usepackage{mathrsfs}
\usepackage{hyperref}
\usepackage[pdftex]{graphicx}
\newcommand{\scs}{\scriptscriptstyle}
\newcommand*{\mydprime}{^{\prime\prime}\mkern-1.2mu}
\newcommand{\smallm}{\scs (\!-\!)}
\newcommand{\smallp}{\scs (\!+\!)}

\newcommand{\smallpm}{\scs (\!\pm\!)}

\begin{document}
\title{Thermo-optic hysteresis with bound states in the continuum}
\author{ D. N. Maksimov$^{1,2,3}$,  A. S. Kostyukov$^{1}$,  A. E. Ershov$^{1,3}$, M. S. Molokeev$^{1,2}$, E. N. Bulgakov$^{2,3}$  and V. S. Gerasimov$^{1,3}$}
\affiliation{$^1$IRC SQC, Siberian Federal University, 660041, Krasnoyarsk, Russia}
\affiliation{$^2$Kirensky Institute of Physics, Federal Research Center KSC SB
RAS, 660036, Krasnoyarsk, Russia}
\affiliation{$^3$Institute of Computational Modelling SB RAS, Krasnoyarsk, 660036, Russia}
\date{\today}
\begin{abstract}
We consider thermo-optic hysteresis in a silicon structure supporting bound state in the continuum. Taking into account
radiative heat transfer as a major cooling mechanism we constructed a non-linear model describing the optical response. It
is shown that the thermo-optic hysteresis can be obtained with low intensities of incident light $I_0\approx 1~\rm{W/m^2}$ at the red edge
of the visible under the critical coupling condition.
\end{abstract}
\maketitle

\section{Introduction}

Recently, we have seen a surge of interest to bound states
in the continuum (BICs)~\cite{Hsu16, Koshelev19, sadreev2021interference} that have grown to an important
tool in nanophotonics paving a
way to optical devices with enhanced light-matter interaction.
BICs do not couple the incident light, however, if the symmetry of the system is broken the
BICs are observed as narrow Fano resonances in the scattering spectrum~\cite{Kim,Shipman,SBR,Blanchard16}.
In more detail, the BICs are spectrally surrounded by a leaky band of high{-}quality resonances (quasi-BICs) which can be excited
from the far-field~\cite{Yuan17}. The excitation of the strong resonances results in critical field enhancement~\cite{Yoon15, Mocella15a}
 with the near-field amplitude controlled by the frequency and the angle of incidence of the
incoming monochromatic wave.
The critical field enhancement triggers nonlinear optical effects even with a low intensity of the
incident light.
This resonant enhancement of nonlinear effects can lead to symmetry breaking~\cite{Bulgakov11},
channel dropping~\cite{Bulgakov13}, excitation of non-linear standing waves~\cite{Yuan16}
as well as self-adaptive robust~\cite{bulgakov2014robust} and tunable Fabry-Perot~\cite{Bulgakov2010} BICs.

In the field of nonlinear optics the BICs have been applied
for second harmonic  generation~\cite{ndangali2013resonant, koshelev2019meta, wang2018large}.
Quasi-BICs in subwavelength dielectric resonators~\cite{rybin2017high, bogdanov2019bound} are also shown
to demonstrate to enhance second harmonic generation~\cite{carletti2018giant, koshelev2020subwavelength}.
In~\cite{yuan2020excitation} it was found that, otherwise decoupled, BICs can be excited via second harmonic generation by illuminating
the structure from the far field.
At the same time it was shown theoretically that quasi-BICs allow for optical bistability~\cite{Yuan17,Krasikov18,Bulgakov19,maksimov2020optical}
due to the Kerr effect.

More recently, the research focus shifted towards BICs in lossy structures. In the presence of material absorption
the BIC are shown to acquire finite-life, albeit remain localized and decoupled from the outgoing channels~\cite{Hu20}.
Quasi-BIC in lossy periodic structures are found to be instrumental for enhancement of light absorption~\cite{Saadabad21, Tan22}
in the critical coupling regime even in low loss dielectrics. This opens novel opportunities for highly
efficient light absorbers~\cite{zhang2015ultrasensitive,wang2020controlling,sang2021highly,xiao2021engineering, cai2022enhancing}.

Lately, it has been suggested that thermo-optical effects can be the dominating nonlinear effects
in BIC supporting structures due to heating by absorbed radiation\cite{zograf2021all}.
In this work we investigate resonantly enhanced thermo-optical bistability~\cite{Sun10, Khandekar17, Gao17a, Pottier21, Ryabov22} in
a system supporting an optical BICs. Taking into account
radiative heat transfer as a major cooling mechanism we shall construct a non-linear model based on the temporal coupled mode theory (TCMT)~\cite{Fan03}
and theoretically demonstrate thermo-optic hysteresis.

\section{Scattering theory}

\begin{figure}
\includegraphics[width=0.49\textwidth, height=0.25\textwidth,trim={0cm 0cm 0cm 0cm},clip]{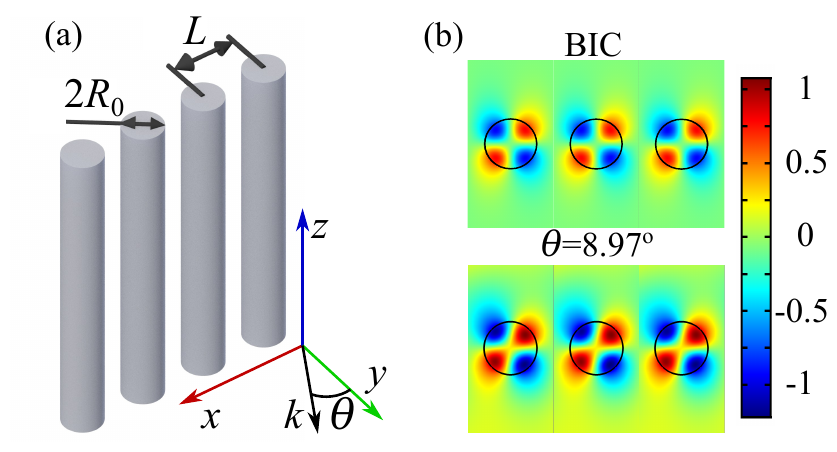}
\caption{(a) Set-up of the array of dielectric rods. (b) Comparison between the eigenmode profile for a BIC with
$\lambda_{\scs{\rm BIC}}=782~\rm{nm}$ and scattering solution under illumination by a plane wave with $\lambda=785~\rm{nm}$.
The solutions are visualized as the $z$-components of the electric field.
The geometric parameters: $R_0=128~{\rm{nm}}$, $L=428~\rm{nm}$.}
\label{fig1}
\end{figure}

We consider an array of identical dielectric rods of radius  $R_0$ linearly arranged with
period $L$ in vacuum. The rods are made of amorphous silicon. The axes of the rods are collinear and aligned with the $z$-axis as shown in Fig.~\ref{fig1}~(a).
Such a system is known to support an abundance of BICs as demonstrated in~\cite{Bulgakov2015,Bulgakov17a}.
In this work we performed numerical simulations with application of FDTD Lumerical to examine the properties of the BIC induced optical response taking
into account both temperature and frequency dependence of the refractive index. In Fig.~\ref{fig1}~(b) we compare the eigenmode profile of an optical BIC with
vacuum wavelength $\lambda_{\scs{\rm BIC}}=782~\rm{nm}$ against the scattering solution obtained under illumination by a TE plane wave with $\lambda=785~\rm{nm}$ at the incidence angle
$\theta=8.97~\mathrm{deg}$, and the incident wave vector on the $x0y$-plane as shown in Fig.~\ref{fig1}~(a). The results are obtained with the numerical values
of the refractive index from~\cite{Jellison94}.
In Fig.~\ref{fig1}~(b) one can see a striking similarity between the two field profiles.

By definition a BIC can not couple
to incident light. However, each BIC is a singular point on the dispersion sheet of a leaky band where the $Q$-factor diverges to infinity. The BIC is, thus,
spectrally surrounded by a family of high-$Q$ leaky modes (quasi-BICs) with the resonant $Q$-factor controlled by variation of the angle of incidence.
In our case the BIC occurs in the $\Gamma$-point being symmetrically mismatched from the single radiation channel of the zeroth diffraction order. Therefore,
the high-$Q$ resonant response is triggered by setting off the angle of incidence from zero. Thus, the similarity between
the BIC and the scattering solution is explained by both BIC and quasi-BIC sitting in the same dispersion band.

Our goal in this section is to set up the optimal regime for enhanced light absorption leading to the most significant change of the refractive index by heating.
The problem of enhanced absorption by quasi-BICs has been previously considered in the literature~\cite{Saadabad21}. The central result for upside-down symmetric
structures, such as the one in Fig.~\ref{fig1}~(a),  is that the maximal absorption of $50\%$ can be achieved in the critical coupling point
where the radiation and absorption loss rates are equal to one another.

The BIC induced optical response can be understood in the framework of temporal coupled mode theory (TCMT)~\cite{Fan03}. For the reader's convenience
we list the most important TCMT formulas below. Let us consider two-channel scattering for TE-polarized light.
The $S$-matrix is implicitly defined
through the following equation
\begin{equation}\label{Scat}
\left(
\begin{array}{c}
s_{1}^{\smallm} \\
s_{2}^{\smallm} \\
\end{array}
\right)
=\widehat{S}
\left(
\begin{array}{c}
s_{1}^{\smallp} \\
s_{2}^{\smallp} \\
\end{array}
\right),
\end{equation}
where $s_{m}^{\smallpm}$ are the amplitudes of the plane waves
in the far-field with subscript $m=1,2$ corresponding
to the upper and lower half-spaces while superscripts $^{\smallp}, ^{\smallm}$ stand for incident and outgoing waves respectively. We assume that the system is illuminated
by a monochromatic wave of frequency $\omega$. In what follows we introduce the vectors of incident and outgoing amplitudes $|s^{\smallpm}(t)\rangle$ which oscillate in time with the harmonic factor $e^{-i\omega t}$.
The TCMT equations~\cite{Fan03} take the following form
\begin{align}\label{CMT_linear}
& \frac{d a(t)}{d t}=-(i\omega_0+\gamma+\gamma_0)a(t)+\langle d^{*}|s^{\smallp}\rangle, \nonumber \\
&|s^{\smallm}\rangle=\widehat{C}|s^{\smallp}\rangle+
a(t)|d\rangle,
\end{align}
where $\widehat{C}$ is the matrix of direct (non-resonant) process, $\omega_0$ is the resonance center frequency,
$\gamma$ is the radiation decay rate, $\gamma_0$ is material loss decay rate, $a$ is the amplitude of the resonant eigenmode, and
$|d\rangle$ is the $2\times1$ vector of coupling constants satisfying
\begin{equation}\label{CMT1}
\langle d|d\rangle=2\gamma.
\end{equation}
The solution
for the $S$-matrix reads
\begin{equation}\label{S}
\widehat{S}=\widehat{C}+\frac{|d\rangle\langle d^*|}{i(\omega_0-\omega)+\gamma+\gamma_0}.
\end{equation}
In the case of the center-plane mirror symmetry
we have
\begin{equation}\label{direct}
\widehat{C}=
e^{i\phi}\left(
\begin{array}{cc}
\rho & i\tau\\
i\tau & \rho
\end{array}
\right),
\end{equation}
where $\phi, \rho, \tau$ are real valued parameters such as
\begin{equation}
\rho^2+\tau^2=1.
\end{equation}
The following equation also holds true~\cite{Fan03}
\begin{equation}\label{CMT2}
\widehat{C}|d^{*}\rangle=-|d\rangle.
\end{equation}
Using Eq.~(\ref{CMT1}) together with Eq.~(\ref{CMT2})  one finds
\begin{equation}
\label{d}
|d\rangle=
e^{i\frac{\phi}{2}}
\sqrt{\frac{\gamma }{2(1+\rho)}}
\left(
\begin{array}{l}
\tau-i(1+\rho) \\
\tau- i(1+\rho)
\end{array}
\right),
\end{equation}
for symmetric modes, and
\begin{equation}
\label{dm}
|d\rangle=
e^{i\frac{\phi}{2}}
\sqrt{\frac{\gamma }{2(1+\rho)}}
\left(
\begin{array}{l}
\tau+i(1+\rho) \\
-\tau- i(1+\rho)
\end{array}
\right),
\end{equation}
for anti-symmetric ones.
 The reflectance can be found from Eq.~(\ref{S}) as
\begin{equation}\label{R}
R=\frac{\rho^2(\omega-\omega_0)^2\mp 2\rho\tau\gamma(\omega-\omega_0)+\tau^2\gamma^2+\rho^2\gamma_0^2}{(\omega-\omega_0)^2+(\gamma+\gamma_0)^2},
\end{equation}
while the transmittance is given by
\begin{equation}\label{T}
T=\frac{\tau^2(\omega-\omega_0)^2\pm 2\rho\tau\gamma(\omega-\omega_0)+\rho^2\gamma^2+\tau^2\gamma_0^2}{(\omega-\omega_0)^2+(\gamma+\gamma_0)^2},
\end{equation}
where the top sign is used for symmetric modes.
Finally, the absorbance is given by
\begin{equation}\label{A}
A=\frac{2\gamma\gamma_0}{(\omega-\omega_0)^2+(\gamma+\gamma_0)^2}.
\end{equation}
The maximal absorbance is obtained at the critical coupling condition
$\gamma=\gamma_0$ and $\omega=\omega_0$. Substituting the above into Eq.~\eqref{A} one finds
\begin{equation}
A_{\mathrm{max}}=\frac{1}{2}.
\end{equation}

\begin{figure}
\includegraphics[width=0.49\textwidth, height=0.25\textwidth,trim={0cm 0cm 0cm 0cm},clip]{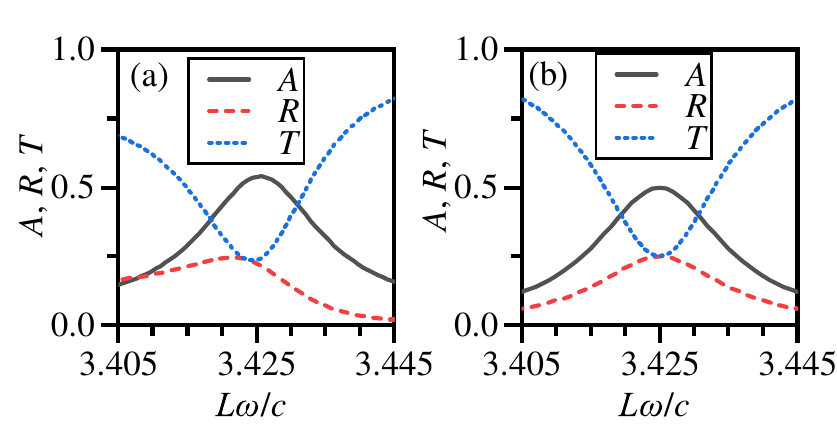}
\caption{Scattering spectra for the system shown in Fig.~\ref{fig1}~(a), $R_0=128~{\rm{nm}}$, $L=428~\rm{nm}$. (a) The reflectance, transmittance, and absorbance spectra found from
FDTD simulations at the critical coupling angle $\theta=8.97~\mathrm{deg}$. (b)
The TCMT approximation of the data in subplot (c); $L\omega/c=3.425$, $L\gamma/c=L\gamma_0/c=0.0056$, $\rho=0$.
}
\label{fig2}
\end{figure}
\begin{figure}[t]
\includegraphics[width=0.49\textwidth, height=0.63\textwidth,trim={0.0cm 0.0cm 0.0cm 0.0cm},clip]{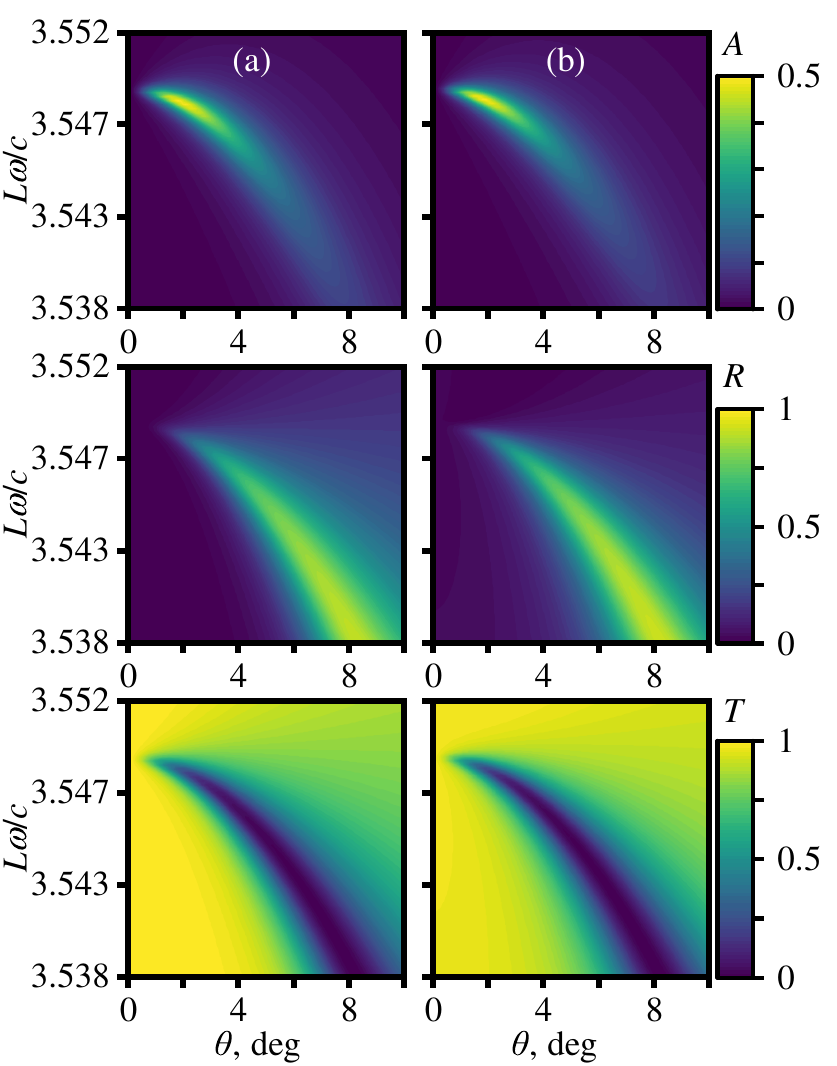}
\caption{Absorbance, reflectance, and transmittance in the spectral vicinity of the BIC in the near infrared, $\lambda_{\scs\rm{BIC}}=1027~\rm{nm}$, $L=580~\mathrm{nm}, R_0=
174~{\rm{nm}}$.
Column (a) shows the data obtained by direct FDTD simulations. Column (b) shows the TCMT approximation, $L\omega_{\scs\mathrm{BIC}}/c=3.549$, $\rho=0$, $L\kappa_{\omega}/c=-0.00016$,
 $L\kappa_{\gamma}/c=0.000072$.  }
\label{fig3}
\end{figure}

Since the system supports its antisymmetric BIC in the $\Gamma$-point, we expect that the resonant frequency $\omega_0$ and the radiation decay rate $\gamma$
are given by the following equations~\cite{Bulgakov17oe}
\begin{align}\label{Taylor1}
& \omega_0=\omega_{\scs \rm {BIC}}+\kappa_{\omega}\theta^2+\mathcal{O}(\theta^4), \nonumber \\
& \gamma=\kappa_{\gamma}\theta^2+\mathcal{O}(\theta^4),
\end{align}
where $\kappa_{\omega,\gamma}$ are the coefficients of the Taylor expansion.
At the same time the absorption decay rate can be assessed in the perturbative manner~\cite{vainstein1988electromagnetic}  as
\begin{equation}\label{gamma0}
\gamma_0=\epsilon\mydprime \int\limits_{S_R} dS \frac{{\bf E}^{\dagger}{\bf E}+{\bf H}^{\dagger}{\bf H}}{4},
\end{equation}
where $\epsilon\mydprime$ and
 ${\bf E}$ and ${\bf H}$ are the electric and magnetic vectors of the BIC eigenfield
satisfying the normalization condition
\begin{equation}\label{normalization}
\int\limits_S dS \frac{\epsilon(x,y){\bf E}^{\dagger}{\bf E}+{\bf H}^{\dagger}{\bf H}}{4}=1,
\end{equation}
with $S$ as  the area of the elementary cell, and $S_R$ as the area of the rod cross section.
By a close examination of Eq.~(\ref{Taylor1}) and Eq.~(\ref{gamma0}) one concludes that the critical coupling
condition can always be achieved by increasing the angle of incidence. In our simulations we found that the critical coupling
condition is fulfilled at $\theta=8.97~\mathrm{deg}$ which is the case shown in Fig.~\ref{fig1}~(b). In Fig.~\ref{fig2}~(a) we show transmittance, reflectance,
and absorbance at the critical angle found via direct numerical simulations. In Fig.~\ref{fig2}~(b) we plotted the TCMT approximation with
Eqs.~(\ref{R},~\ref{T},~\ref{A}). The non-resonant reflection coefficient is taken $\rho=0$ according to~\cite{Bulgakov19}. The other approximation parameters
are listed in the caption to Fig.~\ref{fig2}. By comparing Fig.~\ref{fig2}~(a) against Fig.~\ref{fig2}~(b) one can see a certain deviation from the TCMT approach
which
is due to the perturbative nature of the solution given by Eqs.~(\ref{R},~\ref{T},~\ref{A}). In particular, the absorption in the direct process
is not accounted for by Eq.~\eqref{direct}. A better accuracy of the perturbative solution can be achieved in the near infrared where the imaginary part of
the refractive
index drops by two orders of magnitude~\cite{schinke2015uncertainty}. In Fig.~\ref{fig3} we compare the numerical data for dielectric array with
$L=580~\mathrm{nm}, R_0=
174~{\rm{nm}}$ supporting a BIC at wavelength $\lambda_{\scs\rm{BIC}}=1027~\rm{nm}$. One can see an excellent coincidence between the TCMT and
numerical simulations. Also
notice that the position of the critical coupling spot in the absorption spectrum is shifted towards smaller angles of incidence due to a
smaller $\gamma_0$. Further
on in this work we stay in the visible relying on approximation shown in Fig.~\ref{fig2}~(b), and bearing in mind that the accuracy of the TCMT can be
improved by going
to the near infrared.

\section{Heat transfer}

Since our system consists of dielectric rods in vacuum, the only mechanism of heat transfer is radiation from the rods.
The total energy lost per unit surface area across all wavelengths per unit time is given by the Stefan-Boltzmann law
\begin{equation}\label{SB}
    j=\alpha\sigma[(\Theta_0+\Delta \Theta)^4-\Theta_0^4],
\end{equation}
where $\alpha$ is the emissivity, $\sigma=
5.670\times 10^{-8}
~\mathrm{W}\cdot\mathrm{m}^{-2}\cdot\mathrm{K}^{-4}$ is the Stefan-Boltzmann constant, $\Theta_0$ is the temperature
of the surrounding medium and $\Delta \Theta$ is the temperature increase due to absorption of light. The emissivity, $\alpha$, is a quantity of key importance,
that describes the difference in the radiative properties between
the dielectric rod and the black body. In this work we assessed the emissivity by computing the
absorption coefficient of a single dielectric rod in the temperature range $25^{\circ}\mathrm{C}\leq\Theta\leq 300^{\circ}\mathrm{C}$.
The absorption coefficient was found as the average over all possible direction of the incident wave vector for both polarizations and, then,
averaged with the Planck's distribution
\begin{figure}
    \centering
    \includegraphics[width=0.30\textwidth, height=0.27\textwidth,trim={0cm 0cm 0.0cm 0cm},clip]{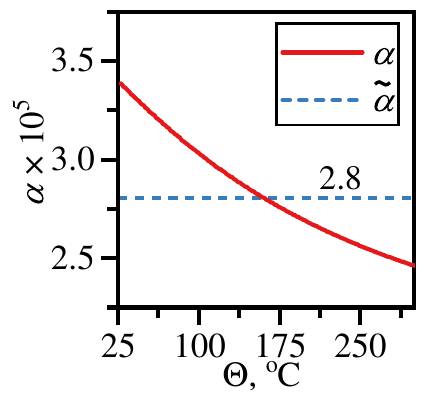}
    \caption{Emissivity of a single dielectric rod as a function of temperature; $L=428$~nm, $R_0=128$~nm. The average emissivity, $\tilde{\alpha}$, is shown
    by the horizontal line.}
    \label{fig4}
\end{figure}
\begin{equation}
    p(\omega)=\frac{\hbar \omega^3}{4 \pi^3 c^2}
    \frac{1}{e^{\hbar \omega/(k_\mathrm{B}\Theta )}-1}
\end{equation}
 peaked at about $10~\mu{\mathrm{m}}$. The numerical values of the refractive index were taken from~\cite{Chandler-Horowitz05}. The temperature dependance of
 the refractive index was neglected in the mid-infrared.
 The result is demonstrated in Fig.~\ref{fig4}. One can see that the emissivity
is dependant on temperature of the rods due to the shift of the peak of Planck's distribution. For further convenience we computed
the temperature average $\tilde{\alpha}=2.8\times10^{-5}$ which will be applied in our simulations. The average emissivity is shown in  Fig.~\ref{fig4} as a horizontal line.

The regime of heating can be determined by the value of the Biot number~\cite{bergman2011fundamentals}
\begin{equation}\label{Bi}
\mathrm{Bi}=\frac{h_S}{k}L_0,
\end{equation}
where $L_0$ is the characteristic length, $h_S$ is the surface heat transfer coefficient, and $k$ is the thermal conductivity of the body.
Values of the Biot number smaller than $0.1$ imply that the heat conduction inside the body is much faster than the heat lost
at the interface between the body and the surrounding medium. In this regime we can assume that the temperature is constant across the body.
By using Eq.~\eqref{SB} we find
\begin{equation}\label{transfer_coeff}
h_S=4\tilde{\alpha}\sigma\Theta_0^3\Delta\Theta.
\end{equation}
Taking $L_0$ as the perimeter of the rod, and $k=1.8~\mathrm{W}/(\mathrm{m}\cdot\mathrm{K})$ according to~\cite{Wada96} we find at the room temperature
$\mathrm{Bi}=6.5\cdot10^{-11}$. This allows us to assume that the temperature is uniform in the cross-section of the rod and avoid simulating heat transfer within the rods. Now our TCMT and heat transfer models can be linked by equating
the energy absorption rate in a unit length of the rod
\begin{equation}\label{energy1}
    \frac{dE}{dt}=\gamma_0(\Theta)|a|^2
\end{equation}
to the radiation rate from the surface per unit length of the same rod
\begin{equation}\label{energy2}
   \frac{dE}{dt}=2\pi R_0 \tilde{\alpha}\sigma[(\Theta_0+\Delta \Theta)^4-\Theta_0^4].
\end{equation}

\begin{figure}[t]
    \centering
    \includegraphics[width=0.37\textwidth, height=0.28\textwidth,trim={0cm 0cm 0.0cm 0cm},clip]{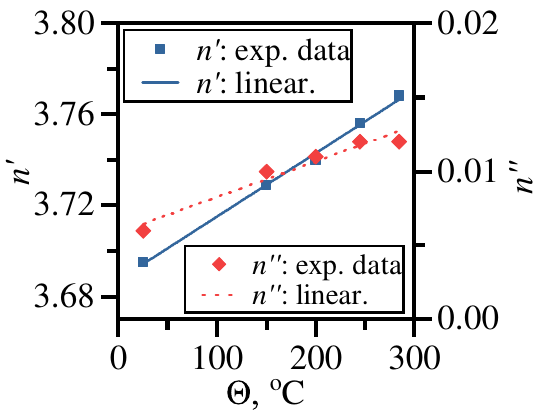}
    \caption{Real $n'$ and imaginary $n''$ parts of the refractive index of the amorphous silicon as a function of temperature at $\lambda=782~\mathrm{nm}$. Solid
    lines show linear approximation.}
    \label{fig5}
\end{figure}
\section{Non-linear TCMT equation and optical bistability}
\begin{figure}[ht]
\includegraphics[width=85mm]{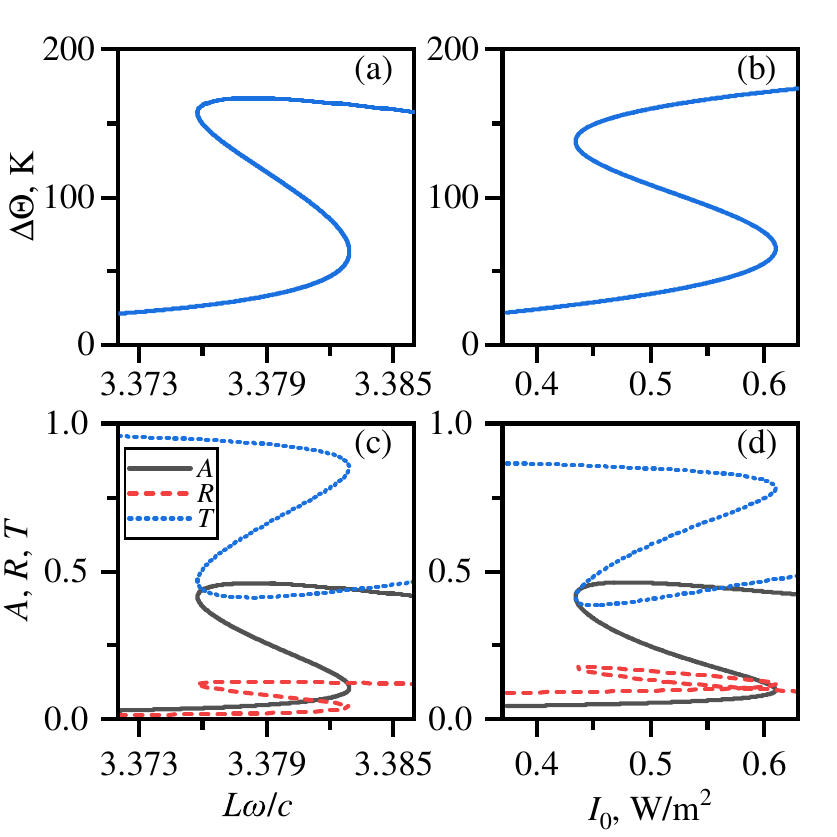}
\caption{Thermo-optic hysteresis in the system shown in Fig.~\ref{fig1}~(a) with the same values of geometric parameters at incident angle $\theta=8.97^{\circ}$.
(a) Temperature increment against frequency of incident light, $I_0=0.54 \rm{W/m^2}$. (c) Scattering spectra, $I_0=0.54 \rm{W/m^2}$.
(b) Temperature increment against the intensity of incident light with vacuum wavelength $\lambda=795.5 \rm{nm}$. (d) Absorbance, reflectance, and transmittance
against the intensity, $\lambda=795.5 \rm{nm}$.}
\label{fig6}
\end{figure}
Frequency domain coupled mode equation reads
\begin{equation}\label{CMT_frequency}
[i\omega_0(\Theta)-i\omega+\gamma(\Theta)+\gamma_0(\Theta)]a=d_1(\Theta)\sqrt{LI_0},
\end{equation}
where $I_0$ is the intensity of the incident monochromatic light. Notice that according to Eq.~\eqref{normalization} the BIC eigenfield is
normalized to store a unit energy in the elementary cell per unit length of the rod. Thus, the intensity in Eq.~\eqref{CMT_frequency}
is multiplied by the period under the basic TCMT assumption
that the scattering channel function are normalized to supply (retract) a unit energy
in a unit of time~\cite{Fan03}.
All parameters including the resonant eigenfrequency, absorption and radiation decay rates are dependent on temperature. Hence we write
\begin{align}\label{Taylor2}
    & \omega_0(\Theta)=\omega_0^{\scs{(0)}}
    +\omega_0^{\scs{(1)}}\Delta\Theta, \nonumber \\
    & \gamma_0(\Theta)=\gamma_0^{\scs{(0)}}
    +\gamma_0^{\scs{(1)}}\Delta\Theta, \nonumber \\
    & \gamma(\Theta)_=\gamma^{\scs{(0)}}
    +\gamma^{\scs{(1)}}\Delta\Theta.
\end{align}
As it was mentioned the temperature dependence is due to the temperature dependance of the refractive index. By using
the experimental data from~\cite{Jellison94} we plotted the temperature dependance of both real $n'$ and imaginary $n''$ parts
of the refractive index of the amorphous silicon at $\lambda=782~\mathrm{nm}$ in Fig.~\ref{fig5}. One can see in Fig.~\ref{fig5} that
both quantities can be linearly approximated in the temperature range of interest.

By numerically computing the scattering spectra with
$\theta=8.97~\mathrm{deg}$ at elevated temperatures
and performing a TCMT fit we found the numerical values of the expansion coefficients in Eq.~\eqref{Taylor2} as follows
\begin{align}
& L\omega_0^{\scs{(1)}}/c=-2.80\times 10^{-4},~\mathrm{K}^{-1}, \nonumber \\
& L\gamma_0^{\scs{(1)}}/c=2.67\times10^{-5},~\mathrm{K}^{-1}, \nonumber \\
& L\gamma^{\scs{(1)}}/c=6.67\times10^{-6},~\mathrm{K}^{-1}.
\end{align}
One can see from the above data that the dominating effect is a shift of the resonant frequency. The resonant frequency is at least one
order of magnitude more sensitive to heating than the other TCMT parameters. Also notice
that the absorption loss rate is much more sensitive to heating than the radiative loss rate. This can be understood from the explicit linear dependance
of $\gamma_0$ on the imaginary part of the refractive index given by Eq.~\eqref{gamma0}. In what follows we neglect
the temperature dependance of $\gamma$ and, consequently, of $|\bf{d}\rangle$.
From \eqref{CMT_frequency} and \eqref{CMT1} we find
\begin{equation}
    |a|^2=\frac{\gamma I_0L}{[\omega-\omega_0(\Theta)]^2
    +[\gamma+\gamma_0(\Theta)]^2},
\end{equation}
which in combination with Eq. \eqref{energy1} and Eq. \eqref{energy2} yields
\begin{align}\label{final}
   & \frac{ (\gamma_0^{\scs{(0)}}
    +\gamma_0^{\scs{(1)}}\Delta \Theta)\gamma I_0L}{(\omega-\omega_0^{\scs{(0)}}
    -\omega_0^{\scs{(1)}}\Delta \Theta)^2
    +(\gamma+\gamma_0^{\scs{(0)}}
    +\gamma_0^{\scs{(1)}}\Delta \Theta)^2}= \nonumber \\
    & 2\pi R_0 \tilde{\alpha}\sigma[(\Theta_0+\Delta \Theta)^4-\Theta_0^4].
\end{align}
The above formula constitutes a non-linear equation which has to be solved for $\Delta\Theta$.

By solving Eq.~\eqref{final} numerically in the frequency domain we find that a bistabily window of width $\Delta\lambda=1.6\rm{nm}$
exists at
\begin{equation}
\frac{2\pi R_0 \tilde{\alpha}\sigma}{I_0L}=6.15\cdot 10^{-12}
\end{equation}
at the incident wavelength $\lambda=795.9~\rm{nm}$. This in combination with our assessment for the emissivity $\tilde{\alpha}=2.80\cdot 10^{-5}$
yields the incident intensity $I_0=0.54~\rm{W/m^2}$. Our findings are illustrated in Fig.~\ref{fig6}. Figure~\ref{fig6}~(a), and Fig.~\ref{fig6}~(b)
demonstrate thermo-optical hysteresis in the frequency domain. One can see that at the peak of the resonance the system is heated up to $\Theta=187~^{\circ}\rm{C}$.
Figure~\ref{fig6}~(c), and Fig.~\ref{fig6}~(d)
demonstrate thermo-optical hysteresis in the intensity domain with the width of bistability window $\Delta I_0=1.6~\rm{W/m^2}$.
\section{Summary and Conclusions}

We considered thermo-optic hysteresis in a system supporting a symmetry protected bound state in the continuum. It is found that
bistability occurs at surprisingly low intensities of incident light, about $I_0=0.54~\rm{W/m^2}$. Two aspects are of key importance for
obtaining low bistability thresholds. (i) Critical coupling which is obtained by breaking the symmetry of the system by incident light.
With the increase of the angle of incidence the radiation decay rate grows quadratically until it becomes equal to the absorption decay rate,
$\gamma=\gamma_0$. In this situation, no matter how small the absorption decay rate is, $50\%$ of the incident light is absorbed and transformed to heat.
This approach paves a way for engineering effective absorbers with low loss dielectrics. (ii) Low emissivity which is due to a small optical path within the absorbing
dielectric at the peak wavelength of Planck's distribution at about $10~\rm{\mu m}$. The dielectric array is assembled of rods with a subwavelength cross section
in the visible. Thus, the optical path in the mid-infrared is at least an order of magnitude less than the wavelength of the thermal radiation. Given that the imaginary part
of the refractive index of silicon is vanishingly small in the mid-infrared~\cite{Chandler-Horowitz05}, $n''\approx 10^{-4}$, this results in a small value of
the emissivity $\tilde{\alpha}=
2.8\cdot 10^{-5}$.

The bistability occurs with intensities eleven orders of magnitude smaller than those needed for hysteresis due to Kerr nonlinearity in silicon
\cite{maksimov2020optical}. Thus, that thermo-optic effects should dominate in non-linear response of silicon nano-structures supporting bound states in
the continuum. In this work we have used a rough estimate for the emissivity. Yet, it is worth mentioning that even with $\tilde{\alpha}=1$, i.e. with
a mid-infrared
black body, the bistability thresholds would drop by less than five orders of magnitude and still be dominating in the non-linear response. Here we neglected
diffusive and convective heat transfer. In particular, the system under consideration has no substrate, which is necessary for realistic dielectric
metasurfaces. One would expect that the presence of substrate would further decrease the bistability thresholds as the heat would be leaking into the substrate.
This issue will be addressed in future studies.

\section*{Acknowledgments} This work received financial support through the grant of Russian Science Foundation and Krasnoyarsk
Regional Fund of Science No~22-22-20056, https://rscf.ru/project/22-22-20056/,
the authors acknowledge discussions with A.A. Bogdanov.

\bibliography{BSC_light_trapping}


\end{document}